\documentclass[cits]{PoS}

\title{Non-perturbative running of the coupling from four flavour 
       lattice QCD with staggered quarks\thanks{IFT-UAM/CSIC-10-59 }}

\ShortTitle{4-flavour running coupling}

\author{\speaker{Paula Perez Rubio,} \thanks{October 2010:  
         Institut f\"ur Theoretische Physik, Universit\"at Regensburg, 
         D-93040 Regensburg, Germany}
        \\
        Instituto de F\'isica Te\'orica UAM-CSIC, 
        28049 Cantoblanco, Spain,\\
        School of Mathematics,Trinity College, Dublin 2, Ireland.\\
        E-mail: \email{Paula.Perez-Rubio@physik.uni-regensburg.de }}

\author{Stefan Sint\\
        School of Mathematics, Trinity College, Dublin 2, Ireland.\\
        E-mail: \email{sint@maths.tcd.ie}}

\abstract{ Using the Schr\"odinger functional (SF) with a single 
           staggered fermion field  we calculate the SF coupling 
           in four-flavour QCD for a wide range of energies and 
           lattice sizes up to $L/a=16$.  Preliminary results for the
           continuum extrapolation of the step-scaling function are 
           presented. To reduce cutoff effects, one-loop ${\rm O}(a)$ 
           improvement has been implemented. Various cross checks are 
           made possible by the use of two independent sets of lattices 
           with either $T=L+a$ or $T=L-a$.}

\FullConference{The XXVIII International Symposium on Lattice Field Theory, 
                Lattice2010\\
		June 14-19, 2010\\
		Villasimius, Italy}

\begin{document}

\section{Introduction}

\vspace{-2mm}



The free parameters of QCD can be fixed by matching a corresponding
number of hadronic observables at low energies. Then, predictions can in
principle be made for any other observable both at low and high
energies. A quantity of particular interest is the strong coupling
constant $\alpha_s$. A commonly used reference value is the
$\overline{\rm MS}$ coupling of dimensional regularisation at the scale
of the $Z$-boson mass.
This poses a particular challenge for the lattice formulation, due to
the large scale differences involved, which cannot be accommodated on
presently affordable lattices. This difficulty can be overcome by using
finite size techniques
\cite{Luscher:1991wu,Luscher:1992zx}. 
The essential ingredient is the
non-perturbative definition of a running coupling, $\bar{g}(L)$, which
runs with the linear extent $L$ of the space-time volume. 
The Schr\"odinger functional provides a framework for such a definition,
%
where $\bar g(L)$ is defined through the response of the system to 
a constant colour electric background field. See 
\cite{Luscher:1992an} for a detailed explanation and 
\cite{Sint:1993un} for the definition of the SF for QCD both 
in the continuum and on the lattice with Wilson quarks.

Within this framework, the step scaling function, (SSF),
plays a fundamental role. It can be regarded 
as a discretised version of the Callan and Symanzik $\beta$ 
function, and can be used to study the evolution of 
$\bar g(L)$ with energy. Such studies have been carried out 
for different numbers of flavours 
\cite{Luscher:1992zx, Luscher:1993gh,DellaMorte:2004bc,
Takeda:2004xha,Aoki:2009tf,Tekin:2010mm}. 
We present here the step scaling function and the 
running coupling with four flavours of staggered quarks. 
In \cite{Tekin:2010mm} an analogous work using 
$\rm{O}(a)$ improved Wilson quarks was presented. 
The agreement of the results in the continuum limit 
can  thus be regarded as a test of universality.

This write-up is organised as follows. Section 2 reviews the basics of 
the SF on the lattice with staggered fermions, 
gives a definition for the renormalised running coupling, and revisits 
the basics on finite size techniques. In section 
3, the ${\rm O}(a)$ improvement for our setup is revisited. Section 4 
presents the details of the simulations and the data analysis. In section 
5, we present our results and we finish with an outlook to future work. 

\vspace{-3mm}

\section{Schr\"odinger functional, coupling constant and finite size 
techniques}

\vspace{-2mm}

The SF is a useful tool to study the 
scaling properties of QCD. Here we will introduce it briefly. See 
\cite{Luscher:1992an,Sint:1993un,Sint:1995rb,Sint:1995ch,Bode:1999sm}
for  more details and \cite{Miyazaki:1994nu, Heller:1997pn, 
PerezRubio:2007qz} for the set up of the SF with staggered fermions. 
It can be regarded as the Euclidean time evolution kernel 
for going from a state at time $x_0= 0$ to another state at time 
$x_0=T$. Using the transfer matrix formalism, it can be expressed 
as a path integral with fields satisfying periodic boundary conditions 
in space and Dirichlet boundary conditions in time. Homogeneous 
boundary conditions are imposed on the fermionic fields, 
\vspace{-2mm}
\be
P_+ \psi\big\vl_{x_0=0} =0 = \bar \psi P_+\big\vl_{x_0=T},
\qquad
P_- \psi\big\vl_{x_0=T} =0 = \bar \psi P_-\big\vl_{x_0=0},
\vspace{-1mm}
\ee
with $P_\pm = \frac12(1\pm \gamma_0)$ and the spatial 
components of the gluon fields satisfy, 
\vspace{-1mm}
\be
A_k(x)\big\vl_{x_0=0}= C_k, \qquad A_k(x)\big \vl_{x_0=T} = C_k'.
\vspace{-1mm}
\ee
The SF is then a functional of these boundary fields, 
\vspace{-1mm}
\be
\mathcal Z[C',C]=\int \mathcal D[A, \psi, \bar\psi]e^{-S[A,\psi, \bar \psi]}.
\vspace{-1mm}
\ee
The choice of the boundary fields is largely arbitrary. Following 
\cite{Luscher:1992an} we choose Abelian and spatially constant fields such 
that the absolute minimum of the action is unique up to gauge 
transformations.  
%
%
%
%
%
%
A judicious choice of the boundary fields ensures that the absolute minimum 
of the action is unique up to gauge transformations and yields small 
lattice artefacts in the renormalised coupling. 

\vspace{-3mm}

\subsection{Definition of the coupling constant} 

The SF  allows us to define a renormalised coupling constant 
$\bar g^2$ non-perturbatively, easily computable on the lattice and 
in perturbation theory and with reasonably
small lattice artifacts. 
Since the induced background field is unique, it is
possible  to unambiguosly 
define the effective action of the SF, i.e., 
\vspace{-3mm}
\be
\Gamma[B] = -\ln\mathcal Z[C, C'].
\vspace{-2mm}
\ee
%
 The boundary fields are parametrised by the scale $L$ 
and  $\eta,\nu $, two dimensionless real parameters
\cite{Luscher:1993gh}. A renormalised coupling can be defined  
through, 
\vspace{-2mm}
\be
\Gamma' = \left.\frac{\partial \Gamma}
{\partial \eta}\right\vl_{\eta=\nu=0}
= \frac{[\partial \Gamma_0/\partial\eta]_{\eta=0}}{\bar g^2(L)},
\vspace{-2mm}
\ee
$\bar g^2(L)$ is going to run with 
the spatial box size $L$ which plays the role of the scale. The  
normalisation constant is chosen so that the renormalised 
coupling coincides with the bare coupling at tree level in the 
perturbative expansion. 

\vspace{-1mm}

\subsection{SF with staggered fermions}

The SF with staggered fermions requires lattices 
where the time extent $T/a$ is odd and the 
spatial extent $L/a$ is even. 
\cite{Miyazaki:1994nu, Heller:1997pn}. This is  
illustrated in figure \ref{fig1} in a two dimensional sketch, as
well as one possible reconstruction of the staggered fermions. See 
\cite{PerezRubio:2008yd} for a more detailed 
description on the fermionic reconstruction. The continuum 
limit for $\bar g^2$ is usually taken setting 
$T=L$ already for finite values of $a$. In order to 
define the continuum limit in our case, we are obliged to 
adopt modified conventions, \cite{PerezRubio:2007qz}. Lattices 
with $T = L\pm a $ are interpreted as having physical time 
extent $T' = T +sa$ with $s = \pm 1$, so that the condition  $T'= L$ 
provides us with 2 regularisations to our problem. This
modification forces us to have the ${\rm O}(a)$ improvement 
revisited.

\begin{center}
\vspace{-3mm}
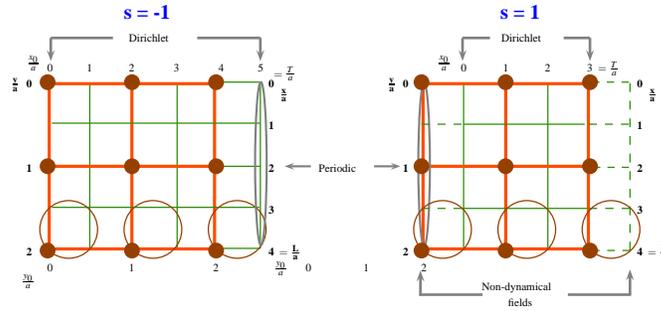
\begin{figure}[ht!]
\begin{center}
\scalebox{0.7} 
{

\begin{pspicture}(0,-2.485)(11.34,2.485)
\definecolor{color38}{rgb}{0.17,0.56,0.027}
\definecolor{color41}{rgb}{0.5,0.5,0.5}
\definecolor{color43}{rgb}{0.59,0.25,0.0}
\definecolor{color45}{rgb}{1.0,0.3,0}

\psline[linewidth=0.024cm,linecolor=color38](0.9,1.56)(0.9,-1.56)
\psline[linewidth=0.024cm,linecolor=color38](2.56,1.56)(2.56,-1.56)
\psline[linewidth=0.024cm,linecolor=color38](3.34,1.6)(4.12,1.6)
\psline[linewidth=0.024cm,linecolor=color38](4.14,1.56)(4.14,-1.56)
\psline[linewidth=0.024cm,linecolor=color38](3.36,-1.56)(4.14,-1.56)
\psline[linewidth=0.024cm,linecolor=color38](4.14,0.0)(3.34,0.0)
\psline[linewidth=0.024cm,linecolor=color38](8,1.56)(8,-1.56)
\psline[linewidth=0.024cm,linecolor=color38](9.6,1.56)(9.6,-1.56)
\psframe[linewidth=0.06,linecolor=color45,dimen=outer](3.3,1.6)(0.1,-1.6)
\psframe[linewidth=0.06,linecolor=color45,dimen=outer](10.4,1.6)(7.2,-1.6)
\psline[linewidth=0.06cm,linecolor=color45](8.8,1.6)(8.8,-1.6)
\psline[linewidth=0.06cm,linecolor=color45](7.2,0.0)(10.4,0.0)
\psline[linewidth=0.06cm,linecolor=color45](1.7,1.6)(1.7,-1.6)
\psline[linewidth=0.06cm,linecolor=color45](0.1,0.0)(3.3,0.0)

\psline[linewidth=0.024cm,linecolor=color38](7.96,-0.8)(10.36,-0.8)
\psline[linewidth=0.024cm,linecolor=color38](0.18,-0.78)(4.12,-0.78)
\psline[linewidth=0.024cm,linecolor=color38](0.18,0.82)(4.14,0.82)
\psellipse[linewidth=0.04,linecolor=color41,dimen=outer](7.25,0.03)(0.13,1.57)
\psline[linewidth=0.024cm,linecolor=color38,linestyle=dashed,dash=0.16cm 0.16cm](7.96,-0.8)(7.2,-0.8)
\psline[linewidth=0.024cm,linecolor=color38,linestyle=dashed,dash=0.16cm 0.16cm](7.96,0.8)(7.2,0.8)
\psline[linewidth=0.024cm,linecolor=color38,linestyle=dashed,dash=0.16cm 0.16cm](10.44,1.6)(11.14,1.6)
\psline[linewidth=0.024cm,linecolor=color38,linestyle=dashed,dash=0.16cm 0.16cm](11.16,1.58)(11.16,-1.6)
\psline[linewidth=0.024cm,linecolor=color38,linestyle=dashed,dash=0.16cm 0.16cm](10.44,-1.6)(11.16,-1.6)
\psline[linewidth=0.024cm,linecolor=color38,linestyle=dashed,dash=0.16cm 0.16cm](10.38,-0.8)(11.16,-0.8)
\psline[linewidth=0.024cm,linecolor=color38,linestyle=dashed,dash=0.16cm 0.16cm](10.46,-0.02)(11.14,-0.02)
\psline[linewidth=0.024cm,linecolor=color38,linestyle=dashed,dash=0.16cm 0.16cm](10.38,0.8)(11.14,0.8)
\psdots[dotsize=0.3,linecolor=color43](0.1,-1.6)
\psdots[dotsize=0.3,linecolor=color43](1.7,-1.6)
\psdots[dotsize=0.3,linecolor=color43](3.3,-1.6)
\psdots[dotsize=0.3,linecolor=color43](0.1,-0.0)
\psdots[dotsize=0.3,linecolor=color43](1.7,-0.0)
\psdots[dotsize=0.3,linecolor=color43](3.3,0.0)
\psdots[dotsize=0.3,linecolor=color43](3.3,1.6)
\psdots[dotsize=0.3,linecolor=color43](1.7,1.6)
\psdots[dotsize=0.3,linecolor=color43](0.1,1.6)
\psdots[dotsize=0.3,linecolor=color43](7.2,-1.6)
\psdots[dotsize=0.3,linecolor=color43](8.8,-1.6)
\psdots[dotsize=0.3,linecolor=color43](10.4,-1.6)
\psdots[dotsize=0.3,linecolor=color43](7.2,-0.0)
\psdots[dotsize=0.3,linecolor=color43](8.8,-0.0)
\psdots[dotsize=0.3,linecolor=color43](10.4,-0.0)
\psdots[dotsize=0.3,linecolor=color43](10.4,1.6)
\psdots[dotsize=0.3,linecolor=color43](8.8,1.6)
\psdots[dotsize=0.3,linecolor=color43](7.2,1.6)
\psline[linewidth=0.024cm,linecolor=color38](8.0,0.8)(10.3,0.8)
\pscircle[linewidth=0.024,linecolor=color43,dimen=outer](0.5,-1.2){0.56}
\pscircle[linewidth=0.024,linecolor=color43,dimen=outer](2.1,-1.2){0.56}
\pscircle[linewidth=0.024,linecolor=color43,dimen=outer](3.7,-1.2){0.56}
\pscircle[linewidth=0.024,linecolor=color43,dimen=outer](7.6,-1.2){0.56}
\pscircle[linewidth=0.024,linecolor=color43,dimen=outer](9.2,-1.2){0.56}
\pscircle[linewidth=0.024,linecolor=color43,dimen=outer](10.8,-1.2){0.56}
\psline[linewidth=0.05cm,linecolor=color41,arrowsize=0.05291667cm 2.0,arrowlength=1.4,arrowinset=0.4]{<-}(4.14,2.0)(4.14,2.44)
\psline[linewidth=0.05cm,linecolor=color41,arrowsize=0.05291667cm 2.0,arrowlength=1.4,arrowinset=0.4]{<-}(0.16,2.0)(0.16,2.44)
\psline[linewidth=0.05cm,linecolor=color41](1.34,2.44)(0.16,2.44)
\psline[linewidth=0.05cm,linecolor=color41](4.14,2.44)(2.94,2.44)
\psline[linewidth=0.05cm,linecolor=color41,arrowsize=0.05291667cm 2.0,arrowlength=1.4,arrowinset=0.4]{<-}(10.4,2.0)(10.4,2.44)
\psline[linewidth=0.05cm,linecolor=color41](10.4,2.44)(9.9,2.44)
\psline[linewidth=0.05cm,linecolor=color41,arrowsize=0.05291667cm 2.0,arrowlength=1.4,arrowinset=0.4]{<-}(7.98,2.0)(7.98,2.44)
\psline[linewidth=0.05cm,linecolor=color41](8.5,2.44)(7.98,2.44)
\psline[linewidth=0.05cm,linecolor=color41,arrowsize=0.05291667cm 2.0,arrowlength=1.4,arrowinset=0.4]{<-}(7.18,-1.98)(7.18,-2.46)
\psline[linewidth=0.05cm,linecolor=color41,arrowsize=0.05291667cm 2.0,arrowlength=1.4,arrowinset=0.4]{<-}(11.16,-1.96)(11.16,-2.44)
\psline[linewidth=0.05cm,linecolor=color41](8.36,-2.44)(7.16,-2.44)
\psline[linewidth=0.05cm,linecolor=color41](11.18,-2.44)(9.98,-2.44)
\psellipse[linewidth=0.04,linecolor=color41,dimen=outer](4.15,0.05)(0.13,1.57)
\psline[linewidth=0.04cm,linecolor=color41,arrowsize=0.05291667cm 2.0,arrowlength=1.4,arrowinset=0.4]{<-}(4.6,-0.0)(5.1,0.0)
\psline[linewidth=0.04cm,linecolor=color41,arrowsize=0.05291667cm 2.0,arrowlength=1.4,arrowinset=0.4]{<-}(6.8,-0.0)(6.3,-0.0)

\put(8.72,2.37){\tiny{Dirichlet}}
\put(1.65,2.37){\tiny{Dirichlet}}
\put(8.35,-2.35){\tiny{Non-dynamical}}
\put(8.85,-2.65){\tiny{fields}}
\put(5.25,-0.1){\tiny{Periodic}}
\put(0.1,1.8){\tiny{0}}
\put(0.85,1.8){\tiny{1}}
\put(1.65,1.8){\tiny{2}}
\put(2.5,1.8){\tiny{3}}
\put(3.35,1.8){\tiny{4}}
\put(4.1,1.8){\tiny{5}}
\put(-0.3,1.9){\tiny{$\frac{x_0}{a}$}}
\put(4.4,1.7){\tiny{$ = \frac{T}{a}$}}
\put(7.5,1.9){\tiny{$  \frac{x_0}{a}$}}
\put(7.95,1.8){\tiny{0}}
\put(8.75,1.8){\tiny{1}}
\put(9.55,1.8){\tiny{2}}
\put(10.35,1.8){\tiny{3}}
\put(10.55,1.8){\tiny{$=\frac{T}{a}$}}
\put(4.5,1.3){\tiny{$\mathbf{\frac{x}{a}}$}}
\put(4.3,1.5){\tiny{\bf 0}}
\put(4.3,0.7){\tiny{\bf 1}}
\put(4.3,-0.1){\tiny{\bf 2}}
\put(4.3,-0.9){\tiny{\bf 3}}
\put(4.3,-1.7){\tiny{\bf 4}}
\put(4.5,-1.7){\tiny{$\mathbf{=\frac{L}{a}} $}}
\put(11.5,1.3){\tiny{$\mathbf{\frac{x}{a}}$}}
\put(11.3,1.5){\tiny{\bf 0}}
\put(11.3,0.7){\tiny{\bf 1}}
\put(11.3,-0.1){\tiny{\bf 2}}
\put(11.3,-0.9){\tiny{\bf 3}}
\put(11.3,-1.7){\tiny{\bf 4}}
\put(11.5,-1.7){\tiny{$\mathbf{=\frac{L}{a}} $}}
\put(1.55,2.8){\textcolor{blue}{\bf{s = -1}}}
\put(8.7,2.8){\textcolor{blue}{\bf{s = 1}}}

\put(4.4, -2.){\tiny{$\frac{y_0}{a}$}}
\put(5.0, -2.){\tiny{0}}
\put(6.1, -2.){\tiny{1}}
\put(7.2, -2.){\tiny{2}}
\put(-0.4, -2.25){\tiny{$\frac{y_0}{a}$}}
\put(0.1, -2.){\tiny{0}}
\put(1.65, -2.){\tiny{1}}
\put(3.25, -2.){\tiny{2}}
\put(-0.6,1.5){\tiny{$\mathbf{\frac{y}{a}}$}}
\put(-0.3,1.5){\tiny{\bf 0}}
\put(-0.3,-0.1){\tiny{\bf 1}}
\put(-0.3,-1.7){\tiny{\bf 2}}
\put(6.55,1.5){\tiny{$\mathbf{\frac{y}{a}}$}}
\put(6.85,1.5){\tiny{\bf 0}}
\put(6.85,-0.1){\tiny{\bf 1}}
\put(6.85,-1.7){\tiny{\bf 2}}

\end{pspicture} 
}
\end{center}
\vspace{-2mm}
\caption{\label{fig1} \footnotesize  
SF with staggered fermions, 2-dimensional sketch. 
Left, $T= L+a$ right, $T= L-a$. Thin (green) lines represent the lattice
where the one-component staggered fermions live, thick (orange) lines
represent the effective lattice for the reconstructed fermions and 
the brown dots stand for the sites where the reconstructed fermions live.}
\vspace{-3mm}
\end{figure}
\vspace{-3mm}
\end{center}

\subsection{Finite size techniques}
 
We want to compute the scale evolution of the renormalised coupling.
We introduce then the step scaling function in the continuum theory, 
\vspace{-2mm}
\be
\sigma(u) = \bar g^2(2L)\big\vl_{u= \bar g^2(L)}.
\vspace{-2mm}
\ee
On the lattice, it can be obtained as the continuum extrapolation 
from a sequence of pairs of lattices with sizes $L/a$ and 
$2L/a$, 
\vspace{-2mm}
\be
\sigma(u) = \lim_{a\to 0} \Sigma(u, a/L).
\vspace{-2mm}
\ee
The procedure is repeated for a range of $u$ values in 
$[\bar g^2(L_{\rm min}), \bar g^2(L_{\rm max})  ]$, 
\vspace{-2mm}
\be
u_0= \bar g^2(L_{\rm min}),\quad u_k= \sigma(u_{k-1})=
\bar g^2(2^kL_{\rm min}), \quad k= 1,2\ldots
\vspace{-2mm}
\ee
After 7, 8 steps, energy differences of ${\rm O}(100)$ 
are  bridged. At sufficiently large energies, perturbation 
theory can be applied to relate the SF coupling $\bar g^2(L)$ with 
a perturbatively defined coupling, e.g. $\bar g^2_{\overline{\rm MS}}$.
At low energies, the connection with physical units can be 
established through the computation of a hadronic quantity, e.g., 
$F_\pi L_{\rm max}$.

\vspace{-4mm}

\section{${\rm O}(a)$ improvement}

\vspace{-3mm}

Due to computer power limitations it is advisable to construct an
action that is ${\rm O}(a)$ improved so that the dominating 
lattice artifacts are cancelled. Following Symanzik's improvement 
programme \cite{Symanzik:1983dc} this can be done by adding 
irrelevant local counterterms to the action, monitored by
adjustable coefficients. These coefficients admit an expansion 
in perturbation theory. In the SF framework additionally to 
the volume counterterms there exist boundary counterterms.

We want our observable, $\Gamma'$ to be ${\rm O}(a)$ improved 
up to one loop in perturbation theory.	The action is taken to be
\vspace{-3mm}
\ba\label{action}
S[U, \chi, \bar\chi] &=& S_g[U] + S_{f}[U,\chi,\bar\chi], \nonumber\\
S_g[U] &=& \frac{1}{g_0^2}\sum_p w(p){\textrm tr }
\left\{ 1 - U(p)\right\},\\
S_{f}[U, \chi,\bar \chi] &=& a^4\sum_{x_0=a}^{T-a}\sum_{{\bf x},\mu}\frac{1}{2a}
\eta_\mu(x)\bar\chi(x)\left[\lambda_\mu U_\mu(x)\chi(x + a\hat\mu )
- \lambda^*_\mu U_\mu (x -a\hat \mu)\chi(x - a\hat \mu)\right], \nonumber
\vspace{-4mm}
\ea
where the sum over the gauge fields runs over all the 
oriented plaquettes $p$  and $w(p)$ are 
weight factors that take the value 1 except for the boundary 
plaquettes, where they take the values $c_t(g_0)$ for the 
time-like plaquettes attached to the boundaries and 
$\frac12 c_s(g_0)$ for the spatial plaquettes. 
Concerning the fermionic part,  $\bar\chi,\chi$ 
represent the one-component fermionic fields,
 homogeneous boundary conditions have been assumed for the 
fermions, a constant phase factor $\lambda_\mu$\footnote{
We have chosen $\lambda_0 =1,\lambda_k = e^{i\theta_k} = 
e^{i\frac \pi 5}$. It leads to a smaller condition 
number of the free fermon matrix \cite{Sint:1995ch}. }  
 has been included, 
and $\eta_\mu= (-1)^{\sum_{\nu \lt \mu}x_\nu/a}$ are the usual 
phase factors of the staggered fermions. 

There are no O($a$) effects arising from the bulk, so that the 
staggered SF will be O($a$) improved by including a couple of boundary 
counterterms. The only pure gauge counterterm relevant in our context 
takes the form, 
\vspace{-2mm}
\be
c^s_t(g_0^2){\textrm tr }\{F_{0k}F_{0k} \}, \quad c^s_t(g_0^2) = c_t^{(0)_s} 
+ g_0^2(c_t^{(1,0)_s} + N_fc_t^{(1,1)_s})+ {\rm O}(g_0^2),
\ee
 where the superscript $s=\pm1$ stands for the 
two possible regularisations. The coefficients above have 
been calculated for our set up obtaining, 
\vspace{-1mm}
\be
c_t^{(0)_s} = \frac{2}{2+s}, \quad
\begin{array}{lll}
c_t^{(0,1)_1} = 0.0274(2), & \quad&c_t^{(1,1)_1} = 0.0077856(4), \\
c_t^{(0,1)_{-1}} =-0.4636(6) , &\quad& c_t^{(1,1)_{-1}} = -0.0266(8). 
\end{array}
\ee
Concerning the fermionic part, there is also a boundary counterterm 
which, in terms of the single-component fields, takes the form 
\cite{PerezRubio:2008yd}, 
\vspace{-2mm}
\ba
\left[d_s(g_0^2) -1\right]\sum_{{\bf x},k}\left\{ \tfrac{1}{2a} \eta_k(x) 
\bar\chi(x)\left[\lambda_k U_k(x)\chi(x+a\hat k) - \lambda_\mu 
U_\mu^\dagger(x-a\hat k)\chi(x-a\hat k) \right] \right\}_{x_0=a}
\nonumber \\ 
\left[d_s(g_0^2) -1\right]\sum_{{\bf x},k}\left\{ \tfrac{1}{2a} \eta_k(x) 
\bar\chi(x)\left[\lambda_k U_k(x)\chi(x+a\hat k) - \lambda_\mu 
U_\mu^\dagger(x-a\hat k)\chi(x-a\hat k) \right] \right\}_{x_0=T-a}
\hspace{-3mm}
\ea
\vspace{-5mm}

\hspace{-7mm}The tree level value of $d_{s}^{(0)}$ was found to be 
$d_{s}^{(0)} = 1+\frac s4$ \cite{Perez-Rubio:PhDthesis}.

\vspace{-2mm}

\section{Simulations and data analysis}

\vspace{-3mm}

In order to carry out our computations, we have made use of a customised 
version of the code offered by
the MILC collaboration \cite{milc:2002}, where the ${\rm O}(a)$ improvement 
presented in the previous section was implemented. The simulations 
have been run for lattice sizes  $L/a = 4,6,8,12,16$ and $s= \pm 1$. 
The statistics range from 60,000 measurements to 160,000 (the quantity 
$\bar{\rm v}$, \cite{Luscher:1992zx} has also been measured). The data 
analysis has been performed by using \texttt{ Uwerr.m} \cite{Wolff:2003sm}.

Instead of tuning the values of $\beta = 2N_c/g_0^2$ for different 
$L/a$ so that they correspond to the same value of the renormalised 
coupling, we have followed the procedure proposed in 
\cite{Appelquist:2009ty}. We measure $\bar g^2(L)$ for a set of 
values $\beta, L/a$ and generate an interpolating function. 
This function is then used to tune $\beta$.  
The interpolation function takes the form, 
\vspace{-2mm}
\be
\frac{1}{\bar g^2(\beta,L/a)} = \frac{\beta}{2N} + \sum_{i=1}^rx_i
\left(\frac{2N}{\beta}\right)^{i-1}.
\vspace{-2mm}
\ee

The data have been fitted by making use of the least squares method. 
The interpolated data inherit two sources of errors, statistical and 
systematic. 

\vspace{-4mm}

\section{Results}

\vspace{-2mm}

\begin{figure}[ht!]

\vspace{-1mm}

\begin{center}

\vspace{-2mm}

\begin{tabular}{cc}
\includegraphics[width=0.4\textwidth]{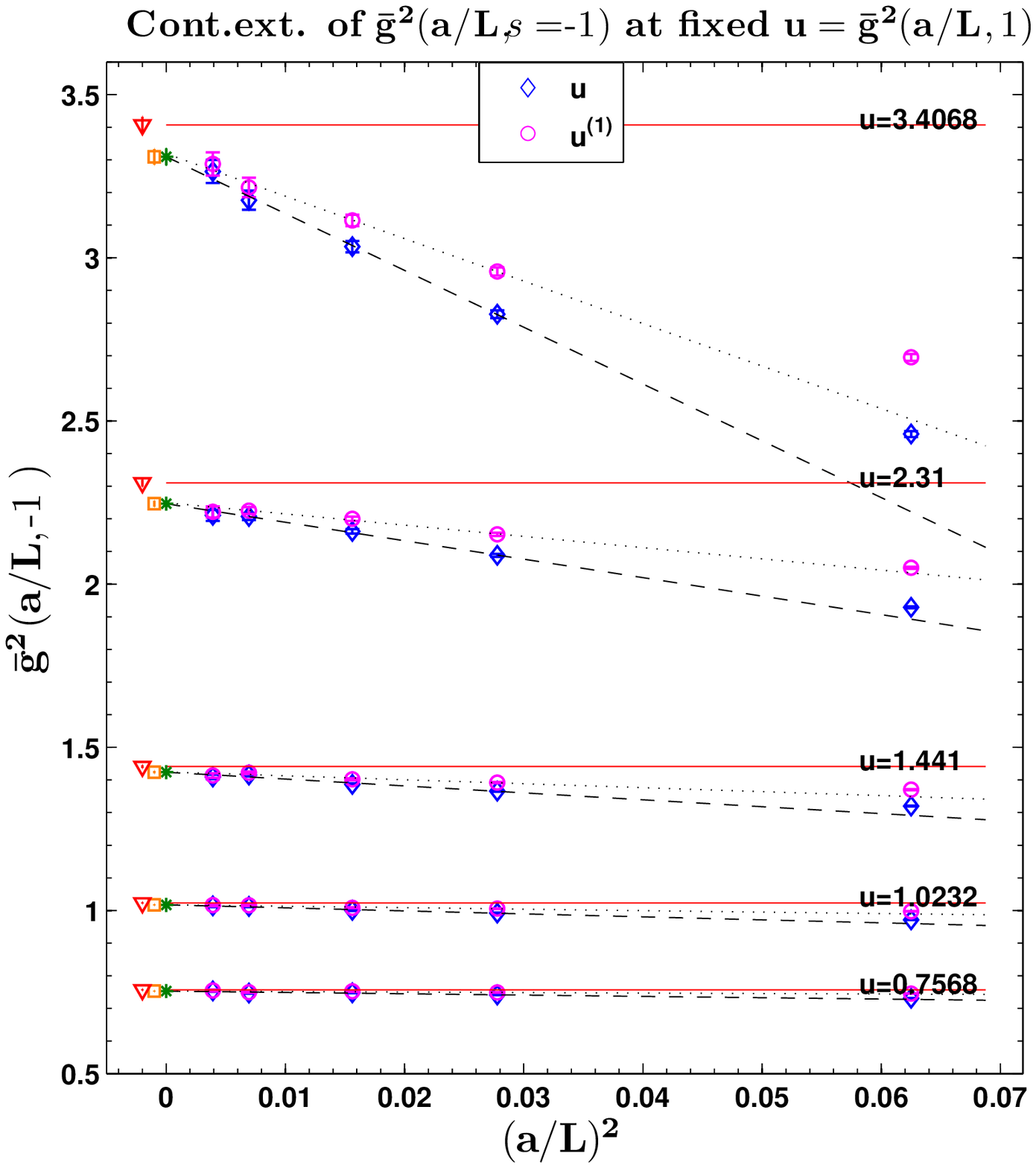}
&
\includegraphics[width=0.4\textwidth]{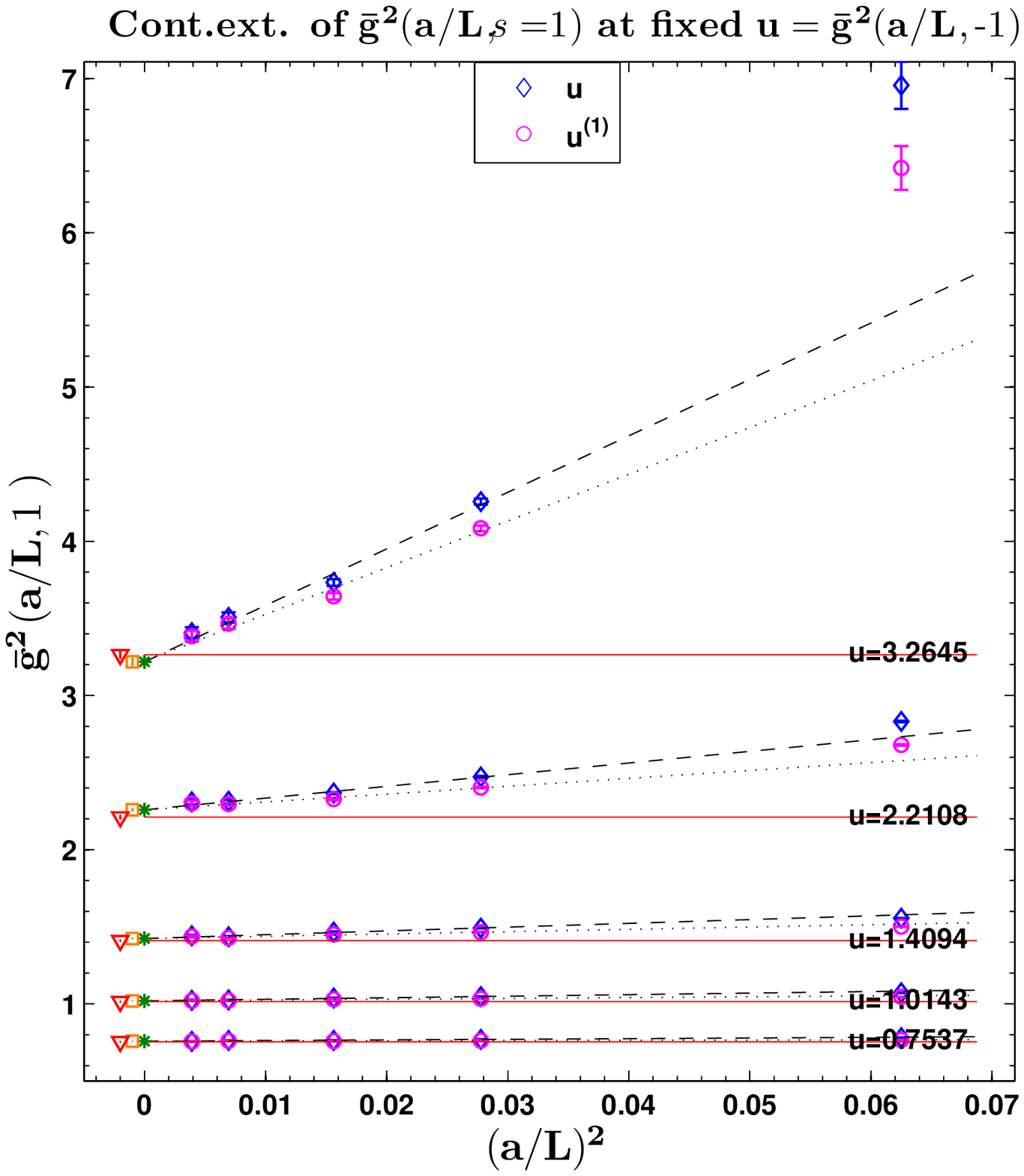}
\end{tabular}

\vspace{-3mm}

\end{center}

\vspace{-7mm}

\caption{\label{fig2} \footnotesize 
Analysis of the lattice artifacts of our system. 
Diamond (blue) points are the values of $\bar g^2(L/a)$. Dashed lines 
correspond to the fit to these data and asterisks (green), the 
continuum extrapolation. Circles, (magenta) are the values for the 
same data after performing the perturbative subtraction, and the dotted 
lines their fits. Squares (orange) represent their continuum 
limit (displaced from the origin). The solid horizontal (red) lines are 
the lines of constant physics, given also by a (red) triangle slightly 
displaced from the origin. $L/a=4$ are not included.}

\vspace{-1mm}

\end{figure}

\begin{figure}[ht!]

\vspace{-2mm}
\begin{center}
\begin{tabular}{cc}
\includegraphics[width=0.4\textwidth]{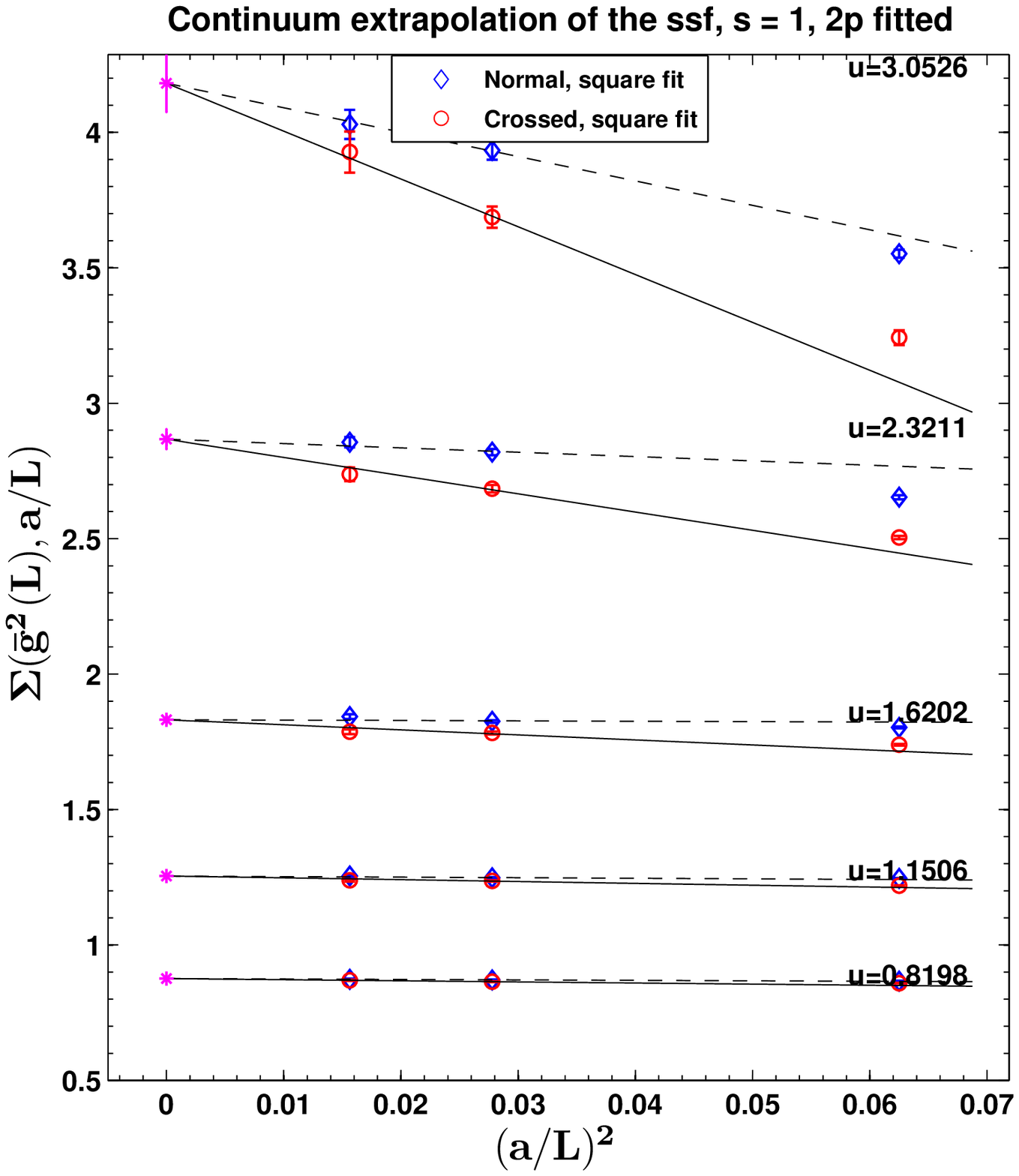}
&
\includegraphics[width=0.4\textwidth]{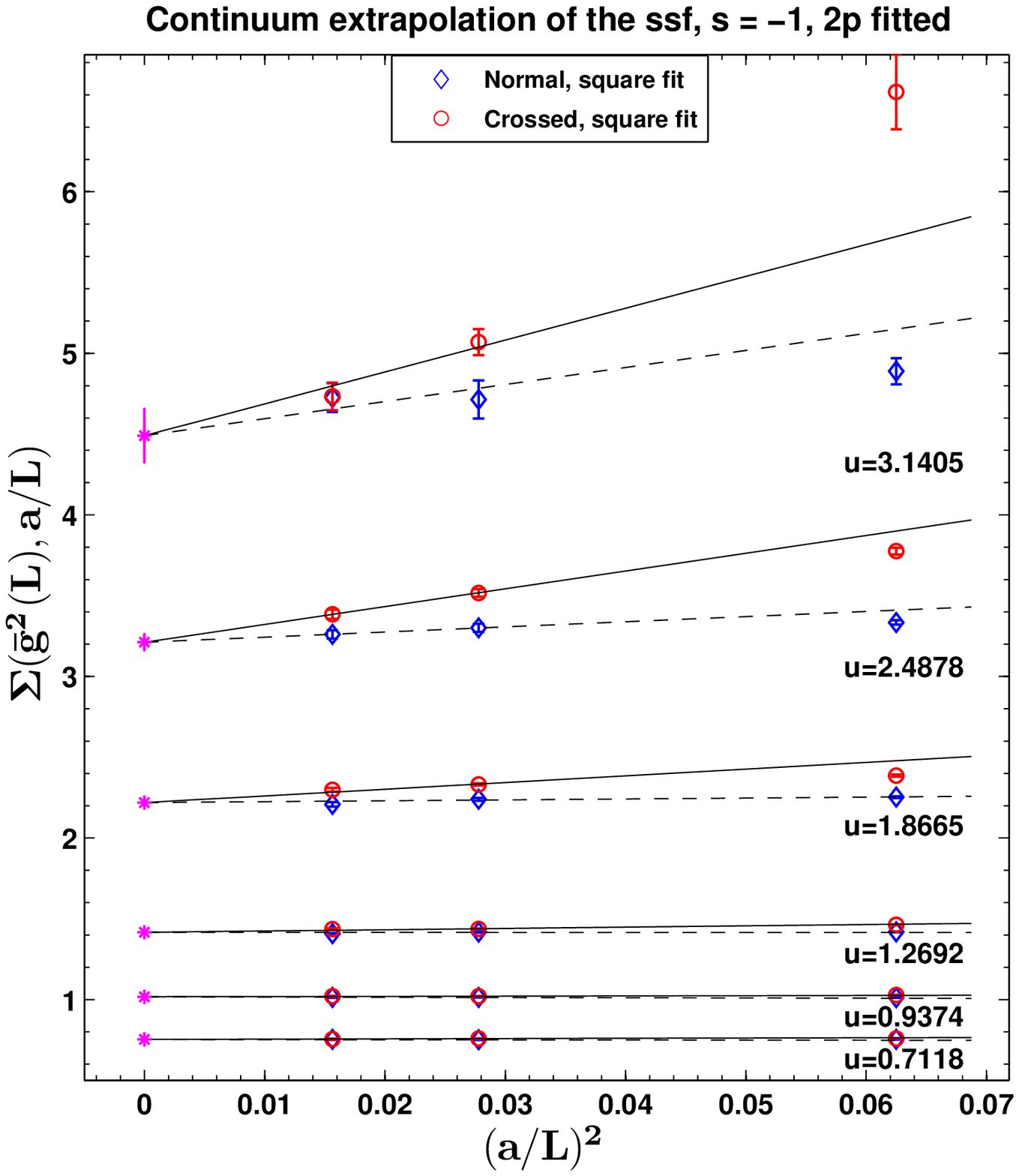}
\end{tabular}
\end{center}
\vspace{-9mm}

\caption{\label{fig3}\footnotesize Continuum limit extrapolation of the step
scaling function. Diamond (blue) points are the values of the step scaling 
function using data from the same regularisation and the dashed lines 
represent the fits. Circles (red) represent the lattice step scaling
function taking data from two regularisations and the solid lines their 
fits. Asterisks (magenta) are the continuum extrapolations $\sigma(u)$ 
of the lattice step scaling function. The renormalisation prescriptions, 
(values of $u$) are explicitly given in the plots. The graph on the right 
correspond to the regularisation $s=1$ and the one on the left to $s=-1$.} 

\vspace{-6mm}

\end{figure}

Since we have two regularisations at our disposal, it is possible 
to perform an analysis of the lattice artifacts of our data. A line
of constant physics is defined by the coupling in 
 one regularisation and we evaluate 
the coupling for the other regularisation at the same values of $\beta$. 
One-loop perturbative cutoff effects can be subtracted, by defining the quantity, 
\vspace{-2mm}
\be
u_\pm^{(1)} =u_\pm\times \left[1+u_\mp(m_1^\pm - m_1^{\mp})\right]^{-1},
\vspace{-2mm}
\ee
where $m_1^\pm$ is the coefficient in the perturbative expansion, 
$\bar g_\pm^2 = g_0^2 + m_1^\pm g_0^4$ and $\pm$ stands for $s= \pm1$
In Figure \ref{fig2} we show the lattice artifacts for $u$ and $u^{(1)}$.
We have also measured the continuum extrapolation of the step scaling 
function. If we choose one regularisation to fix the physics, the lattice 
step scaling function can be computed in the two available regularisations.
The continuum limit has to be shared. Perturbative effects can be 
subtracted,  by defining $\Sigma^{(1)}_s$, 
\vspace{-1mm}
\be
\Sigma^{(1)}(u,a/L) = \frac{\Sigma(u,a/L)}{1+ \delta_1(a/L)u},\quad 
 \delta  = \tfrac {\Sigma(u,a/L) - \sigma(u)}{\sigma(u)}= 
\delta_1(a/L) u + \delta_2(a/L)u^2 +{\rm O}(u^3).
\vspace{-2mm}
\ee
 In figure \ref{fig3} we can see the  continuum extrapolation 
for some of our data. The fit has been done in the following way, 
\vspace{-2mm}
\be
\Sigma_1(u_s, a/L) =\sigma_{\rm mixed}(u_s) + A_1(a/L)^2,\quad
\Sigma_{-1}(u_s, a/L) =\sigma_{\rm mixed}(u_s) + A_2(a/L)^2, 
\vspace{-2mm}
\ee
where $s$ stands for the regularisation  chosen 
to fix the physics. The fit was performed excluding the 
data for  $L/a=4$. An analogous fit was done for 
$\Sigma^{(1)}_{\pm1}(u_s,a/L)$. 

\begin{figure}[ht!]
\vspace{-3mm}

\begin{center}
\includegraphics[width=0.5\textwidth]{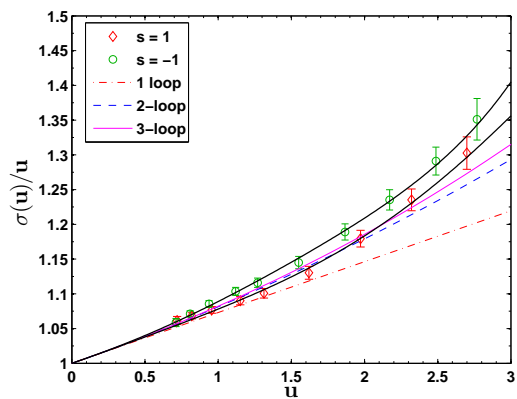}

\vspace{-5mm}

\caption{\label{fig4}\footnotesize
 Step scaling function $\sigma(u)$. The dotted-dashed 
(red), dashed (blue) and solid (magenta) lines represent 
the perturbative 1-loop, 2-loop and 3-loop $\sigma(u)$. Diamonds (red) 
represent the extrapolated $\sigma(u)$ from the $s=1$ regularisation 
and circles (green) from the $s=-1$. Their fits (largest value excluded) are given by the solid thick (black) lines.} 
\vspace{-3mm}
\end{center}

\vspace{-5mm}

\end{figure}

We interpolate the data of the continuum extrapolation of $\sigma(u)$. 
In Figure \ref{fig4} we present such interpolation, plotting $\sigma(u)/u$
vs $u$, together with the perturbative approximations to 1,2 and 3 
loop in PT. Our fit function was a polynomial of degree 6 where 
the first coefficients where set to the perturbative coefficients up to 
2 loop \cite{Bode:1999sm} 
in PT. The values corresponding to the largest couplings were not included 
in the fits. In the plot, we show the results obtained from the two 
different regularisations. They are correlated  and therefore 
they can not be simultaneously used in a fit. The differences 
between the two can be regarded as systematic errors. 
\vspace{-3mm}

\vspace{-2.5mm}

\section{Conclusions and outlook}

\vspace{-3mm}

We have computed the SSF of the QCD coupling in the 
SF scheme with 4 flavours of massless staggered quarks.
Unfortunately, the discretisation errors are  fairly large. Some 
more effort will  be put into the data analysis, 
and the  $\Lambda$ parameter still needs to be computed. 
The results are in rough agreement with data obtained with Wilson quarks 
\cite{Tekin:2010mm}, but a detailed comparison is still needed.

\vspace{-2mm}

\section*{Aknowledgements}

\vspace{-2mm}

This research was supported by the Research Executive Agency (REA) of the
European Union under Grant Agreement number 
PITN-GA-2009-238353 (ITN STRONGnet), by the Ministerio de Educaci\'on de 
Espa\~na through an FPU grant, the Universidad Aut\'onoma de Madrid 
and Trinity College Dublin. 
We are grateful for the 
support of the Trinity Centre for High-Performance Computing (TCHPC), and
the Irish Centre for High End Computing (ICHEC)  where the simulations 
were carried out.

\end{document}